\documentclass{PoS}

\usepackage{amsmath}
\usepackage{simplewick}

\def\ep{\varepsilon}
\def\half{\frac{1}{2}}

\def\K{\mathcal{K}}
\def\C{\mathcal{C}}
\def\A{\mathcal{A}}

\def\C{\mathcal{C}}
\def\K{\mathcal{K}}
\def\A{\mathcal{A}}
\def\I{\mathcal{I}}
\def\F{\mathcal{F}}

\def \bear {\begin{eqnarray}}
\def \eear {\end{eqnarray}}
\def\lr#1{\left(#1\right)}
\def\slr#1{\left[#1\right]}

\def\trl#1{\textrm{Tr}\lr{#1}}

\def\avg#1{\left\langle #1\right\rangle}

\def\CPn{\mathbb C P^n}

\def\RS2{\mathbb R\times S^2_F}

\def\Xint#1{\mathchoice
   {\XXint\displaystyle\textstyle{#1}}%
   {\XXint\textstyle\scriptstyle{#1}}%
   {\XXint\scriptstyle\scriptscriptstyle{#1}}%
   {\XXint\scriptscriptstyle\scriptscriptstyle{#1}}%
   \!\int}
\def\XXint#1#2#3{{\setbox0=\hbox{$#1{#2#3}{\int}$}
     \vcenter{\hbox{$#2#3$}}\kern-.5\wd0}}

\def\dashint{\Xint-}

\def\bse{\begin{subequations}}
\def\ese{\end{subequations}}

\def \be  {\begin{equation}}
\def \ee  {\end{equation}}
\def \bex  {\begin{equation*}}
\def \eex  {\end{equation*}}
\def \bea {\begin{eqnarray}}
\def \bal {\begin{align}}
\def \eea {\end{align}}

\def\no{\nonumber\\}

\def\pd#1#2{\frac{\partial #1}{\partial #2}}

\def\reff{{r_{eff}}}

\title{Fuzzy field theories and related matrix models}

\ShortTitle{Fuzzy field theories and related matrix models}

\author{M\'aria \v Subjakov\'a\\
        Department of theoretical physics,\\ Faculty of mathematics, physics and informatics, Comenius University,\\
        Mlynska Dolina,  842 48 Bratislava, Slovakia\\
        E-mail: \email{maria.subjakova@fmph.uniba.sk}}

\author{\speaker{Juraj Tekel}\\
        Department of theoretical physics,\\ Faculty of mathematics, physics and informatics, Comenius University,\\
        Mlynska Dolina,  842 48 Bratislava, Slovakia\\
        E-mail: \email{juraj.tekel@fmph.uniba.sk}}

\abstract{We review the description of scalar field theories on fuzzy spaces by Hermitian random matrix models. After reminding the reader of the relevant aspects of the random matrix theory and construction of the fuzzy spaces, we summarize the most important results for the scalar fields on such spaces. We then introduce the multi-trace matrix models relevant for the analytical description of scalar field theories on fuzzy spaces and show to what extent they do, and to what extent they do not, describe the know phase structure of $\phi^4$ theory on the fuzzy sphere.}

\FullConference{Corfu Summer Institute 2019 "School and Workshops on Elementary Particle Physics and Gravity" (CORFU2019)\\
		31 August - 25 September 2019\\
		Corfù, Greece}

\begin{document}

\section{Introduction}

This review summarizes the description of fuzzy field theories in terms of Hermitian random matrix models.

Noncommutative spaces have first been considered as a regularization to tame the infinities of quantum field theories \cite{snyder}, but soon after that renormalization proved to be very efficient in doing so and the rest is history. Later, it has been realized that quantizing gravity should lead to a spacetime with some nontrivial structure at short distances \cite{doplicher}. Noncommutative spaces provide such structure in a way that preserves (at least some) of its symmetries \cite{sF22}, but it was soon realized that this very feature also brings the phenomenon of UV/IR-mixing \cite{uvir1,uvir2}, spoiling renormalizability. Therefore, field theories on noncommutative spaces cannot be used as fundamental description of nature.

In this text, we will consider a special case of noncommutative spaces, which are as classical manifold compact and come with the name fuzzy spaces \cite{madore0}. This name reflects their short distance structure and inability to locate things completely precisely. They arise in effective descriptions of condensed matter phenomena \cite{qhe2} and as solutions and backgrounds in various matrix formulations in string theory \cite{br1,br2}. Here, we will consider them as toy models of spaces with short distance structure.

The prime example of such spaces is the fuzzy sphere \cite{sF22,madore}, a finite mode approximation to the classical two dimensional sphere. We will construct this object and describe some physics on it in the section \ref{sec3}. Afterwards, in the section \ref{sec4}, we will analyze scalar field theory on the fuzzy sphere using techniques of random matrix models. However, to make this review self-contained at least to certain extent, and to introduce our conventions and notation, let us start with a section describing random matrix theories.

\section{Tools of random matrix theory}

We start by introducing the technical tools we will need. These are the results of the Hermitian random matrix theory. We will first describe the basic concepts, then introduce our main technical approach - the saddle point approximation, give some explicit formulae for our needs and finally generalize the standard results to the setting important for the models describing scalar field theory on fuzzy spaces, i.e. the multi-trace matrix models.

We will be perhaps little brief in our presentation, as our main interest lies further ahead. For more details, we refer the interested readers to the excellent reviews \cite{saclay,matrixmodels,matrixth,2dqg} or to some of the original papers \cite{brezin,japonec,newcritical}.

\subsection{Random matrix ensembles}

Random matrix ensemble is exactly what the name suggests. A matrix with random entries determined by certain probability distribution. We will not describe the topic in its whole generality and we will concentrate on the ensemble most relevant for our needs. This is the ensemble of the $N\times N$ Hermitian matrices $M$. In such ensemble, the expected value of a matrix function $f(M)$ is given by
\begin{align}
\avg{f}\,=\,&\frac{1}{Z}\int dM\,e^{-N^2S(M)}f(M)\ ,\label{3.1}\\
	&dM=\prod_{i=1}^n dM_{ii}\prod_{i<j} d\textrm{Re}M_{ij}\,d\textrm{Im}M_{ij}\,\nonumber 
\end{align}
where $Z$ is normalization such that $\avg 1=1$. Our probability measure is determined by the function $S(M)$, which is often referred to as an action, and we will deal with probability measures of the following form\footnote{The quadratic coefficient is usually written like $g_2=r/2$.}
\be\label{singletraceaction}
S(M)=\frac{1}{N}\sum_{n=0}^N g_n \trl{M^n}=\frac{1}{N}\trl{V(M)}\ .
\ee
The importance of the various factors of $N$ will become clear as we move along. The motivation for their introduction is to accommodate the large $N$ limit, which we are eventually going to be interested in. We can scale the quantities of the model, i.e. the matrix $M$ and the couplings $g_n$, with $N$. We use this to ensure that all the terms in the probability measure behave the same way in the large $N$ limit and thus all contribute to the final result and are not scaled away.

The important property of such expression is its invariance under conjugation of the matrix $M$ with a unitary matrix $M\to UMU^\dagger$, where $U\in SU(N)$. Since also the measure $dM$ is invariant under such conjugation, in such ensembles both the matrix $M$ and $UMU^\dagger$ have the same probability. We can go a step further and rewrite the average (\ref{3.1}) in terms of new variables, the eigenvalues of the matrix $M$ and the unitary matrix $U$ that mixes them\footnote{This $U$ is not unique, but the number of such matrices is the same for every configuration of eigenvalues and thus contributes just as an overall factor.}
\begin{align}
    M=U\Lambda U^\dagger\ ,\ \Lambda=\textrm{diag}(\lambda_1,\ldots,\lambda_N)\ .
\end{align}
In terms of these new parameters the measure becomes
\begin{align}
    dM=\lr{\prod_{i<j}(\lambda_i-\lambda_j)^2}\lr{\prod_{i=1}^N d\lambda_i} dU\ ,\label{3.7}
\end{align}
where the Jacobian, which turns out to be the square of the Vandermonde determinant, can be computed in the vicinity of the identity matrix, since the measure is $U$ invariant, and follows from
\be
dM_{ij}=d\lambda_i\delta_{ij}+(\lambda_i-\lambda_j)dU_{ij}\ .
\ee

\subsection{Saddle point approximation}
The expectation values (\ref{3.1}) we are interested in became, for functions that depend only on the eigenvalues of the matrix,
\begin{align}\label{sadlef}
	\avg{f}=\frac{1}{Z}\int \lr{\prod_{i=1}^N d\lambda_i}&f(\lambda_i)\,e^{-N^2\slr{\frac{1}{N}\sum_i V(\lambda_i)-\frac{2}{N^2}\sum_{i<j}\log|\lambda_i-\lambda_j|}}\times\int dU\ ,
\end{align}
where we have raised the Vandermonde determinant to the exponent. The $dU$ integration is an overall factor and gets absorbed into the normalization.

We will now consider the large $N$ limit of this expression. In this limit, something interesting happens. Both sums in the square brackets are of order $1$ for eigenvalues finite in the large $N$ limit\footnote{This can be achieved through scaling of the matrix $M$ with $N$.}. The $N^2$ factor in the exponent then ensures that only configurations of eigenvalues which evaluate to a small number in the square brackets contribute significantly to the integral. As we increase the value $N$, fewer and fewer configurations are relevant for the evaluation of the integral and in the $N\to\infty$ limit the whole integral localizes onto a configuration which minimizes the expression in the brackets. Notice that the choice of $N^2$ scaling in (\ref{3.1}) was chosen to match the $N^2$ scaling of the Vandermonde term.

This condition translates into least-action-like conditions
\be
\pd{S}{\lambda_i}=0\ \ \ \forall i\ .
\ee
This straightforwardly leads to the following saddle point condition
\be\label{2saddle}
V'(\lambda_i)-\frac{2}{N}\sum_{j\neq i}\frac{1}{\lambda_i-\lambda_j}=0\ \ \ \forall i\ .
\ee
There is a very surprising connection between the saddle point configuration and classical dynamics of point particles. As we can see, the condition (\ref{2saddle}) is an equilibrium condition for $N$ particles in an external potential $V(\lambda)$ with a logarithmic repulsion between each pair of particles. Intuition from such a particle picture will help us greatly to understand the solutions of the saddle point conditions.

To solve (\ref{2saddle}), we introduce the resolvent function
\be \omega(z)=\frac{1}{N}\sum_i\frac{1}{z-\lambda_i}=\int d\lambda\frac{\rho(\lambda)}{z-\lambda}\ ,\label{defomg}\ee
where $\lambda_i$ are from now on the solutions of the large $N$ conditions (\ref{2saddle}), rather than the random numbers in (\ref{sadlef}). We have introduced the distribution of the eigenvalues 
\be
\rho(\lambda)=\frac{1}{N}\sum_j \delta(\lambda- \lambda_j)\ ,
\ee
which we expect to converge to some function in the large $N$ limit. Notice that the resolvent is well defined for any complex $z$ not in the support of the density $\rho(\lambda)$. Using the Sokhotski–Plemelj formula the condition (\ref{2saddle}) becomes 
\be \omega(z+i\ep)+\omega(z-i\ep)=V'(z)\label{3.25}\ee
and expression for the density follows also
\be \rho(\lambda)=-\frac{1}{2\pi i}\slr{\omega(\lambda+i\ep)-\omega(\lambda-i\ep)}\ .\label{3.2525}\ee
We have managed to turn the saddle point problem (\ref{2saddle}) into a Rieman-Hilbert like problem (\ref{3.25}) for the function $\omega(z)$. This can be attacked head on using the machinery of complex analysis \cite{complex}, but for the particular form of the problem we can obtain the solution more easily.

Computing the square of the resolvent, dropping a term subleading in the large $N$ limit and using the saddle point condition we get a quadratic equation 
\be
\omega(z)^2=V'(z)\omega(z)-P(z)\ ,
\ee
where $P(z)$ is a polynomial yet to be determined. This has a straightforward solution
\be\
\omega(z)=\half\slr{V'(z)-\sqrt{V'(z)^2-4P(z)}}\ .
\ee
Support of the eigenvalue distribution is given by the polynomial in the square root, which has to be of an even order. Simple real roots of the polynomial will determine the endpoints of intervals, which give this support. Let us write this part as a product of $(z-z_i)$ for all simple real roots $z_i$ and denote it $\sigma(z)$. The rest needs to be positive, so it does not contribute to discontinuity of, and thus needs to be in the form $M(z)^2$ for some polynomial with no simple roots in the intervals defined by $z_i$.

We thus write 
\be\label{22solresol}
\omega(z)=\half\slr{V'(z)-|M(z)|\sqrt{\sigma(z)}}\ .
\ee
Note that this means that there needs to be an even number of the roots. The same form of $\omega(z)$ can be deduced directly from (\ref{3.25}) after recalling the discontinuity properties of the square root. Finally (\ref{3.2525}) leads to
\be \rho(\lambda)=\frac{1}{2\pi i}|M(z)|\sqrt{\sigma(z)}\ .\ee
What remains to be determined is the form of the two polynomials $M(z)$ and $\sigma(z)$. The latter determines the support of the density $\rho(\lambda)$ and we need to go over all the possible types of support. The former is then given by the condition
\be \omega(z)\to\frac{1}{z}\ \ \ \textrm{for }z\to\infty\ ,\label{omgcond}\ee
which is a result of the definition (\ref{defomg}) and the normalization of $\rho(\lambda)$. This means that any polynomial terms from $V'(z)$ in (\ref{22solresol}) need to cancel with polynomial terms from the second term.

Let us stress that once we know the distribution of the eigenvalues $\rho(\lambda)$, all the large $N$ expectation values can be computed using
\be \avg{f}=\int d\lambda\rho(\lambda)f(\lambda)\ .\ee

There might be more than one solution to the saddle point condition (\ref{2saddle}). If more $\rho$'s are possible, we need to use the most probable one in the expression above. This means the preferred solution is the one with the lowest exponent in the probability distribution (\ref{sadlef}). In the particle dynamics analogy, this corresponds to the inverse temperature $\beta=1/N^2$ being lowered as we take $N\to\infty$, which leads to the particles finding the global minimum of their energy even though there might be several local minima available.

This quantity is called the free energy and is given by\footnote{Free energies defined in this way are finite in the large $N$ limit. Similar expressions are sometimes defined such that $\F$ is of order $N^2$.}
\begin{align}
\F\,=\,&\slr{\frac{1}{N}\sum_i V(\lambda_i)-\frac{2}{N^2}\sum_{i<j}\log|\lambda_i-\lambda_j|}=\no=\,&\slr{\int d\lambda\, V(\lambda)\rho(\lambda)-\dashint  d\lambda d\sigma\, \rho(\lambda)\rho(\sigma)\log|\lambda - \sigma|}\ .
\end{align}
The double integral in this expression is often complicated to calculate, but there is a nice trick, which eliminates this need. We start with the action that needs to be minimized and introduce a condition of normalization of $\rho(\lambda)$
\begin{align}
S_V(\rho)\,=\,&\int d\lambda\, V(\lambda)\rho(\lambda)-\dashint  d\lambda d\sigma\, \rho(\lambda)\rho(\sigma)\log|\lambda - \sigma|-\xi\lr{1-\int d\lambda \, \rho(\lambda)}\ .
\end{align}
where we have added a subscript $V$ to stress the difference from the original action in the probability measure. Variation $\delta S_V/\delta \rho$ leads to the condition
\be V(\lambda)-2\dashint d\sigma \rho(\sigma)\log|\lambda -\sigma|+\xi=0\ .\ee
Note that derivative of this expression with respect to $\lambda$ yields the saddle point condition (\ref{2saddle}). The condition holds for any value of $\lambda$ and can be used to compute the Lagrange multiplier $\xi$. When $\xi$ is known we get
\be \F=N^2\slr{\half\int d\lambda\, V(\lambda)\rho(\lambda)-\half\xi}\ .\ee

To conclude this section, let us note that the moments $c_n$ of the distribution $\rho(\lambda)$ are straightforwardly accessible from the expansion of resolvent function as
\be
\omega(z)=
\frac{1}{z}\sum_{n=0}^\infty \frac{1}{z^n}\int d\lambda\,\lambda^n\rho(\lambda)=
\frac{1}{z}\sum_{n=0}^\infty \frac{c_n}{z^n}\ .
\ee

\paragraph{One-cut assumption.}

We make the simplest assumption on the support of the distribution $\rho(\lambda)$, namely that it is supported on a single interval $(a,b)$. Since such interval represents a cut in the complex plane, where the resolvent function $\omega(z)$ is not defined, this kind of solution is referred to as one-cut solution. This means that $\sigma(z)=(z-a)(z-b)$ and we get\footnote{$\textrm{Pol}$ denotes the polynomial part, i.e. the part of the large $z$ expansion that does not vanish in the $z\to\infty$ limit.}
\be\label{22polyres}
\textrm{Pol}\,\slr{V'(z)-M(z)\sqrt{(z-a)(z-b)}}=0\ .
\ee
We have dropped the absolute value and we will remember that if $M(z)$ becomes negative in the support of $\rho(\lambda)$, the whole solution is inconsistent with one-cut assumption and we need to start over again. The polynomial $M(z)$ is given by
\be\label{22polyM}
M(z)=\textrm{Pol}\,\frac{V'(z)}{\sqrt{(z-a)(z-b)}}\ ,
\ee
and the endpoints are given by the condition that there is no constant term in the large $z$ expansion of $\omega(z)$ and that the leading order is precisely $1/z$.

\paragraph{Two-cut and higher cut assumption.}

The next simplest assumption on the support of the distribution $\rho(\lambda)$ is of the two interval nature. This means
\be \sigma(z)=(z-a)(z-b)(z-c)(z-d)\ .\ee
The eigenvalue distribution is given by conditions similar to (\ref{22polyres}) and (\ref{22polyM}). Similarly for higher number of cuts. What however can, and will, happen is that there are not enough conditions to determine the endpoints of the intervals and there is a whole family of certain classes of solutions.

If the potential is polynomial of degree $s$, there are $s+1$ conditions from (\ref{omgcond}), but for the $n$-cut solution we need to fix $2n$ conditions. 

This makes perfects sense from the particle dynamics point of view. Intervals in the support of $\rho(\lambda)$ are located around the minima of potential $V(\lambda)$. In a stable situation, particles are going to be spread around these minima due to the mutual repulsion. However sometimes, taking one or more eigenvalues from one minimum to the other does not destabilize the situation. It leads to a different stable situation and thus the solution is not given uniquely in that case.

\subsection{Quartic potential}\label{sec2quart}
We will present complete results for the quartic potential case \cite{brezin,japonec}
\be\label{22actionquartic}
S(M)=\half r\, \trl{M^2}+g\,\trl{M^4}\ .
\ee
The reason is twofold. This is the simplest case that illustrates all the features of the Hermitian matrix models and will be the relevant case when we will describe the $\phi^4$ scalar field theory on fuzzy spaces.

\subsubsection{One-cut solution}
Since the potential is symmetric, we expect a symmetric solution and we thus set $b=-a=\sqrt \delta$.\footnote{Spoiler: this will turn out to be a hasty assumption.} The condition (\ref{22polyres}) now explicitly reads
\be \textrm{Pol}\,\slr{r z+4g z^3 -M(z)\sqrt{z^2-\delta^2}}=0\ .\ee
This determines the form and the coefficients of $M(z)$ in terms of $r,g$ and $\delta$, which is determined by the condition (\ref{omgcond}). That explicitly reads
\be\label{22quartcond1}
	\frac{3}{4} \delta^2 g + \frac{1}{4}\delta r=1\ \ \Rightarrow\ \  \delta=\frac{1}{6g}\lr{\sqrt{r^2+48g}-r} \ .
\ee
And finally we arrive at the eigenvalue distribution
\be
	\rho(\lambda)=\frac{1}{2\pi}\lr{r+2g \delta+4g\lambda^2}\sqrt{\delta-\lambda^2}\ , \ \lambda^2<\delta.
	\label{3.38}
\ee
The limit $g\to 0$ of this solution recovers the famous Wigner semicircle distribution for the Gaussian unitary ensemble.\footnote{Note that the distribution (\ref{3.38}) exists also for $g<0$ for positive $r$. Such solutions are important when the matrix integral is used to generate random discretization of 2D surfaces \cite{2dqg}.}

There is however a catch. This distribution can become negative for certain values of $r$ and $g$. But we mentioned above this is inconsistent with the one-cut assumption and in such case we need to start our analysis from scratch. After plugging in the formula, we find out that (\ref{3.38}) is a well-defined distribution for
\be r>-4\sqrt{g}\ .\label{ex1c}\ee

Looking at (\ref{22quartcond1}) we see that as we increase the coupling $g$, the interval shrinks. This is quite understandable from the particle dynamics point of view, the mutual repulsion cannot push the particles that far up the steeper walls of the potential. Also, for negative $r$ the force due to potential close to the origin pushes the particles away and thus the density of the particles decreases. For $r$ large enough the repulsion is not strong enough to push the particles up the maximum at $\lambda=0$ and they split into two parts.

For further reference, we give explicit formulae for the free energy
\be
\F_{S1C}=\frac{-r^2 \delta^2 + 40 r \delta}{384} - \half\log\lr{\frac{\delta}{4}} + \frac{3}{8}\ =-\frac{\delta^2 g}{4}-\frac{3 \delta^4 g^2}{128}+\frac{3}{4} - \half\log\lr{\frac{\delta}{4}}.\label{frees1c}
\ee
and the moments of the distribution
\be
c_1=0\ ,\ c_2=\frac{\delta}{4}+\frac{\delta^3 g}{16}\ .
\ee
The more enthusiastic of the readers are encouraged to derive these expressions as an exercise. We have eliminated the parameter $r$ in favor of other parameters, which will turn out to be useful later.

\subsubsection{Two-cut solution}

The second possibility is having a distribution supported on two intervals. We will again assume symmetry of the solution and parametrize the interval as
\be
\C=\lr{-\sqrt{D+\delta},-\sqrt{D-\delta}}\cup\lr{\sqrt{D-\delta},\sqrt{D+\delta}}
\ee
and arrive at the following two conditions
\be
4Dg+r=0\ ,\ \delta^2=\frac{1}{g}
\ee
Note that we require $D>\delta$, or else the solution makes little sense. This yields
\be r<-4\sqrt{g}\ee
which is precisely complementary condition to the existence of the one-cut solution (\ref{ex1c}).

The eigenvalue distribution becomes
\be
	\rho(\lambda)=\frac{2 g |\lambda|}{\pi}\sqrt{\delta^2-(D-\lambda^2)^2}\ ,
\ee
with the free energy
\be 
\F_{S2C}=-\frac{r^2}{16g} + \frac{1}{4}\log\lr{4g} +\frac{3}{8}=-D^2 g\ + \frac{1}{4}\log\lr{4g} +\frac{3}{8}\label{frees2c}
\ee
and moments $c_1=0,c_2=D$.

\subsubsection{Phase transition}\label{sec2phase}

As we have seen above, the solution to the model changes behavior along the line
\be\label{22basictrln}
r=-4\sqrt g\ .
\ee
To see what kind of change this is, we recall a little bit of statistical physics, reintroduce inverse temperature into the game by
\be r\to \beta r\ , \ g\to\beta g \label{simpletrafo}\ee
and compute the derivatives of the free energies (\ref{frees1c}) and (\ref{frees2c}). It is not difficult to see that the free energy itself as well as the first derivative of the free energy with respect to $\beta$ are smooth across the line (\ref{simpletrafo}). However the second derivative
\be C_V=-\left.\frac{\partial^2 \F}{\partial \beta^2}\right|_{\beta=1}=\left\{
\begin{array}{rl}
\frac{1}{4}+\frac{r^4}{3456 g^2}+\frac{r \left(24 g  -r^2 \right)\sqrt{48 g+r^2}}{3456 g^2} & \textrm{ for }r>-4\sqrt{g}\\
\frac{1}{4} & \textrm{ for } r<-4\sqrt{g}
\end{array}
\right.
\ee
is not smooth. This means that the change of behavior of the model is a phase transition proper with the third derivative of the free energy discontinuous. This phase transition is often referred to as the matrix phase transition.

\subsubsection{Asymmetric solutions}\label{sec234}

As we have already mentioned the assumption on symmetry of the support of the distribution $\rho(\lambda)$ has made the life a lot more simple for us, but did hide some solutions to the saddle point equation (\ref{2saddle}) from us. We will gather the asymmetric solutions here.

\paragraph{Asymmetric one-cut solution.} We start with the one-cut solution. Using the following parametrization of the interval
\be
\C=\lr{D-\sqrt\delta,D+\sqrt\delta}\label{ass1c}
\ee
and being very careful with the analysis of the conditions (\ref{omgcond}) and (\ref{22polyres}) we obtain the following two conditions
\be
D\lr{2 D^2 g + 3 \delta g + \half r}=0\ ,\ 3 D^2 \delta g + \frac{3}{4} \delta^2 g + \frac{1}{4}\delta r=1\ .
\ee
There is the $D=0$ symmetric solution (\ref{3.38}) but also an asymmetric solution
\be
\delta=\frac{-r-\sqrt{-60g+r^2}}{15 g}\ ,\ D=\pm\sqrt{\frac{-3r+2\sqrt{-60g+r^2}}{20 g}}\ . 
\ee
This solution exists only for $r<-2\sqrt{15 g}$ and the reader is invited to exercise the particle dynamics analogy to make sense of these formulas.

The distribution itself becomes
\be
\rho(\lambda)=\frac{1}{2\pi}\lr{4 D^2 g+ 4 D g \lambda + 2 \delta g + r  + 4 g \lambda^2}\sqrt{\delta-\lr{\lambda-D}^2}\ ,
\ee
with moments
\be
c_1=3 D^3 \delta g + \frac{3}{2} D \delta^2 g + \frac{1}{4}D \delta r\ ,\ 
c_2=3 D^4 \delta g + 3 D^2 \delta^2 g + \frac{1}{4}\delta^3 g +  \frac{1}{4} D^2 \delta r + \frac{1}{16}\delta^2 r
\ee
and lengthy free energy
\begin{align}
\F_{A1C}\,=\,&\frac{\delta}{8}\Bigg[\frac{9}{16}g^2 \delta^3+\delta^2\lr{\frac{5}{8}gr+\frac{27}{2}g^2D^2}
+\delta\lr{33g^2D^4+\frac{15}{2}g r D^2+\frac{r^2}{8}+\frac{3}{2}g}\no&
+12 g^2 D^6+7gr D^4+\lr{12g+\half r^2}D^2+ r\Bigg]+\frac{1}{4}r D^2+\half g D^4-\half \log\lr{\frac{\delta}{4}}\ .
\end{align}
One can check that this free energy is always greater than the free energy (\ref{frees2c}) of the two-cut solution. Thus even though the asymmetric solution does exist, it is never the preferred solution to the model, again something that is quite intuitive from the particle dynamics point of view.

\paragraph{Asymmetric two-cut solutions.} As was the case for one-cut solution, the symmetry assumption made life easy of us also in the two-cut case but did miss relevant solutions to the model. We will not go through explicit calculations here, which can be found e.g. in \cite{jt18}. The lesson to be learned is that for any value of $r<-4\sqrt{g}$ there are asymmetric two-cut solutions possible, which is quite understandable from the particle dynamics point of view. But all these solutions have higher energy than the symmetric solution and thus do not change the overall picture of the solutions to the model.

But one needs to be not too hasty with conclusions regarding the symmetry of the action under $M\to-M$ to be the reason for this. As we will see, in the section \ref{sec4.2}, one can have symmetric action that has a preferred asymmetric solution.

\subsubsection{Asymmetric quartic potential}

One is naturally led to consider the case of an asymmetric potential
\be
S(M)=a\trl{M}+\half r\, \trl{M^2}+g\,\trl{M^4}\ ,
\ee
which leads to stable asymmetric solutions. We will not deal much more with this issue here, as it is surprisingly complicated and is covered at length in \cite{jt18}, together with the discussion of the two-cut solutions. We will only give explicit conditions for the one-cut solution (\ref{ass1c})
\begin{align}
0\,=\,&\frac{1}{2}a+2 D^3g+3 D \delta g+\half D r\label{3det1}\ ,\\
1\,=\,&3 D^2 \delta g+\frac{3}{4} \delta^2 g+\frac{1}{4}\delta r\ .\label{3det2}
\end{align}
This illustrates the complications, as we arrive at a condition for $D$ which is not solvable analytically and are left with numerical approach only.

The expressions for the distribution $\rho(\lambda)$, the free energy $\F$ and the moments are the same as for the asymmetric solution in the symmetric potential of the section \ref{sec234}, just with modified solutions for $D$ and $\delta$.

\subsection{Multi-trace matrix models}\label{sec2multi}

The form of the action (\ref{singletraceaction}) is a little too restrictive. An action containing products of traces is still invariant under $SU(N)$ conjugation, which was a crucial aspects of our previous analysis. The most general action we can write is
\be S(M)=f(c_1,c_2,\ldots)+\half r \trl{M^2}+g \trl{M^4}\ ,\label{24genmlti}\ee
where the moments are related to the traces by
\be c_n=\frac{1}{N}\trl{M^n}\ .\ee
Such actions containing products of traces are referred to as multi-trace actions, opposing to single-trace models (\ref{singletraceaction}).

Using the saddle point method, which means we work in the large $N$ limit, leads to the following condition \cite{jt15b}
\begin{align}
\sum_n \pd{f}{c_n} n \lambda_i^{n-1}+r \lambda_i+4 g \lambda_i\,=\,&\frac{2}{N}\sum_{i\neq j}\frac{1}{\lambda_i-\lambda_j}\ .
\end{align}
As we can see the function $f$ generates further terms on the potential side of the equation. Multi-traces thus act as an additional interaction among the eigenvalues, but not of a two-particle nature. They give an interaction with an overall configuration of the eigenvalues through the moments of their distribution. The saddle point condition can be rewritten as
\begin{align}
r_{eff}\lambda_i+\sum_{n\neq 2} g_{n,eff}\lambda_i^{n-1}\,=\,&\frac{2}{N}\sum_{i\neq j}\frac{1}{\lambda_i-\lambda_j}\ ,
\end{align}
where the effective coupling constants are functions of the moments of the distribution. This condition is then to be solved using the method of single-trace models described in section \ref{sec2quart}. The resulting distribution is thus not given uniquely, but is formally function of its moments. The equations for the moments of the distribution are then to be viewed as self-consistency conditions that finally determine the distribution.

The free energy is then determined by
\begin{align}
\F\,=\,&f(c_1,c_2,\ldots)+\half r c_2+g c_4-\frac{2}{N^2}\sum_{i<j}\log|\lambda_i-\lambda_j|\ .
\end{align}
It is most convenient to express the sum in this expression using the free energy of the effective model, as we will demonstrate in the explicit calculation for a simple second moment multi-trace model.

\paragraph{An example.} Let the probability distribution be given by the action
\be\label{24simplecorrection}
S(M)=f(c_2)+\half r\trl{M^2}+g\trl{M^4}\ ,
\ee
which yields the effective quadratic coupling
\be \reff=r+2f'(c_2)\ ,\label{24verysimplereff}\ee
Condition (\ref{22quartcond1}) becomes
\be \frac{3}{4} \delta^2 g + \frac{1}{4}\delta r+\half \delta f'\lr{\frac{\delta}{4}+\frac{\delta^3 g}{16}}=1\ .\label{fconditon}\ee
This condition cannot be solved analytically even for the simplest functions $f$ and one is left with a numerical solution.

To obtain the free energy, we use the expression for the free energy of the effective single-trace model
\begin{align}
    \F\,=\,&f(c_2)+\half r c_2+g c_4-\frac{2}{N^2}\sum_{i<j}\log|\lambda_i-\lambda_j|=\no=\,&
    f(c_2)+\half r c_2+g c_4 +\F_{eff}-\half r_{eff} c_2-g_{eff} c_4=\no=\,&f\lr{c_2}-f'(c_2)c_2-\frac{\delta^2 g}{4}-\frac{3 \delta^4 g^2}{128}+\frac{3}{4} - \half\log\lr{\frac{\delta}{4}}=\no=\,&
    f\lr{\frac{\delta}{4}+\frac{\delta^3 g}{16}}+\frac{1}{4}+\frac{9 \delta^4 g^2}{128}+\frac{\delta r}{8}+\frac{1}{32} \delta^3 g r - \half\log\lr{\frac{\delta}{4}}\ .
\end{align}
In the last step, we have used (\ref{fconditon}) to eliminated the $f'$ term and the lack of $r$ in the expression (\ref{frees1c}) is appreciated.

The analysis of the two-cut case yields conditions that are more manageable
\be 4Dg+2f'(D)+r=0\ , \ \delta=\frac{1}{\sqrt{g}}\ .\label{fconditon2}\ee
We obtain the free energy in a very similar fashion
\begin{align} 
\F_{S2C}\,=\,&f(D)-f'(D)D -D^2 g\ + \frac{1}{4}\log\lr{4g} +\frac{3}{8}=\no
=\,&f(D)+D^2 g + \half r D+ \frac{1}{4}\log\lr{4g} +\frac{3}{8}\ .
\end{align}

The phase transition condition, given by $r_{eff}=-4\sqrt{g_{eff}}$, becomes
\be\label{multimatrixtrafo}
r=-4\sqrt{g}-2f'\lr{\frac{1}{\sqrt{g}}}\ .
\ee

However one needs to be very careful for the cases where $f$ is negative. Such functions introduce repulsion among the eigenvalues and can potentially destabilize otherwise stable solution.

\subsubsection{Phase transitions in multi-trace models}
An obvious question is what happens to the phase transition in the multi-trace case. We invoke the same machinery as in the section \ref{sec2phase}
\be r\to \beta r\ , \ g\to\beta g \ ,\ f\to \beta f \ .\ee
One again needs to calculate the derivatives of $\F$ with respect to $\beta$ and look for discontinuities. This is quite a tedious task in general, but it is straightforward to check that in the simplest case $f=c_2^2$ the second derivative of the free energy becomes discontinuous.

In general, multi-trace terms in the probability distribution lower the order of the matrix phase transition of the model.

\section{Fuzzy spaces and fuzzy field theory}\label{sec3}

After introductory discussion of the basic matrix models techniques let us proceed to a more physical, and central, topic of our discussion. The fuzzy spaces and their physics. In this section, we will describe the construction of fuzzy spaces, then define the scalar field theory on fuzzy spaces and give an overview of the most interesting properties of such theories.

As always, there are several great resources for readers looking for more details \cite{noncom1,noncom2,bal,lecturesydri,steinacker_review0,steinacker_review}.

\subsection{Construction of fuzzy spaces}

We will describe the construction of fuzzy spaces giving some detail for the case of the fuzzy sphere and also some ideas regarding a more general fuzzy space.

\paragraph{An appetizer.} Recall that the central property of the fuzzy spaces has been a short distance structure. We will show how one possible way of introducing minimal distance on the sphere naturally leads to noncommutative algebra of functions. The starting point is the decomposition of the functions on the regular sphere into spherical harmonics
\be f(\theta,\phi)=\sum_{l=0}^\infty\ \sum_{m=-l}^l c_{lm}Y_{lm}(\theta,\phi)\ , \  
\Delta Y_{lm}(\theta,\phi)=l(l+1) Y_{lm}(\theta,\phi)\ .\label{regspherefunctions}\ee
This is true for any reasonable function on the sphere and even less reasonable objects, like Dirac $\delta$-function describing a single point, can be expressed as limits of such expressions.

Functions with finer short-distance features require harmonic functions with higher $l$ in their expansion, with a truly fine example, the $\delta$-function, requiring all the harmonics $l\to\infty$. A natural way to limit the possible resolution of features of objects on the sphere is to introduce a cutoff $L$ on the possible angular momentum in (\ref{regspherefunctions}) and consider objects of the form
\be f_L(\theta,\phi)=\sum_{l=0}^L \sum_{m=-l}^l c_{lm}Y_{lm}(\theta,\phi)\ .\label{fuzzyfunctions}\ee
We however quickly run into trouble, since a product of two such objects cannot be expressed again in the form (\ref{fuzzyfunctions}), but in general requires $l$ beyond $L$.\footnote{For example easily seen considering the product $Y_{10}Y_{11}$.} But there is a way out.

The number of spherical harmonics with $l\leq L$ is $(L+1)^2$, which is the same as the number of independent ${(L+1)\times (L+1)}$ Hermitian matrices. Following this lead, we consider a ${N\times N}$ Hermitian matrix as a product of two $N$-dimensional representations $\underline N$ of the group $SU(2)$. This product is a reducible representation and reduces to
\be \begin{array}{rrcccccccc}
	\underline N\otimes \underline N & 
	=&
	\underline 1&
	\oplus &
	\underline 3&
	\oplus& 
	\underline 5&
	\oplus&\ldots\\ 
	&
	&
	\downarrow&
	&
	\downarrow&
	&
	\downarrow&
	&&
	\\	
	&=&\{Y_{0m}\}&\oplus &\{Y_{1m}\}&\oplus &\{Y_{2m}\}&\oplus&\ldots	
	\end{array}\ .\ee	
We thus have a map $\varphi:Y_{lm}\to M$, which maps the basis of objects defined by (\ref{fuzzyfunctions}) onto the basis of the Hermitian matrices. Using $\varphi$, we can define a brand new product between the spherical harmonics
\be Y_{lm}\ast Y_{l'm'}:=\varphi^{-1}\lr{\varphi\lr{Y_{lm}}\varphi\lr{Y_{l'm'}}}\ .\label{fuzzystar}\ee
Objects (\ref{fuzzyfunctions}) with product $\ast$ form a closed (matrix) algebra. This algebra can be viewed as a finite mode approximation of the regular sphere with a short distance structure, while in the limit $N\to\infty$ we recover the standard sphere.

Note that this approximate sphere enjoys the full rotational symmetry of the original sphere, since the functions in (\ref{fuzzyfunctions}) can still be rotated by a rotation of the spherical harmonics in (\ref{fuzzystar}).
	
\paragraph{The fuzzy sphere.}
More formally we can define the the fuzzy sphere $S^2_F$ as an object with the algebra of functions generated by $X_1,X_2,X_3$ satisfying the following conditions
\be \sum_{i=1}^3 X_i X_i = \rho^2 \ \ \ , \ \ \ X_i X_j - X_j X_i=i \theta \ep_{ijk} X_k\ .\ee
These are clearly a generalization of conditions on coordinates of the classical sphere. These can be realized as the ${N=2j+1}$ dimensional representation of the $SU(2)$ by
\be X_i=\frac{2R}{\sqrt{N^2-1}} L_i \ \ \ , \ \ \ \theta=\frac{2R}{\sqrt{N^2-1}}\sim\frac{1}{N} \ \ \ , \ \ \ \rho^2=\frac{4R^2}{N^2-1} j(j+1)=R^2\ ,\label{matrxisphere}\ee
where $L_i$ are the generators in that representation. Note that in the limit $N\to\infty$ we recover the original sphere, since $\theta\to0$. The $X_i$'s generate Hermitian matrices so we have, again, obtained the algebra of Hermitian matrices $M$ as the algebra of real functions on the fuzzy sphere.

We would now like to define other relevant notions for the fuzzy sphere such as derivative and integral. Integral serves as a scalar product on the space of matrices, i.e. will be a trace. With a proper normalization we have
\be \int_{S^2} f\ \to \ \frac{4\pi R^2}{N}\trl{M}\ .\label{fuzzyint}\ee
We thus see that the sphere has been divided into $N$ cells and interpret the eigenvalues of the matrix $M$ as values of the fuzzy function on these cells. Derivative is a little more difficult to obtain, since the proper derivation relies on the symplectic structure on the sphere \cite{steinacker_review0}, but it should come as no surprise that commutator with $X_i$ in (\ref{matrxisphere}) is going to be involved. The spherical Laplacian becomes
\be \Delta\ \to\ \frac{1}{R^2}\sum_{i=1}^3[L_i,[L_i,\cdot]]\ ,\label{fuzzyder}\ee
where from now on we will not explicitly write summations over repeated index. Notice how the above formula for the fuzzy Laplacian obeys the $SU(2)$ symmetry, it is the quadratic Casimir operator in the relevant representation.

\paragraph{More general fuzzy spaces.} Before we proceed to investigate physical consequences of the fuzzy structure, let us briefly mention how the above construction generalizes to other spaces \cite{steinacker_review0,steinacker_review}.

We start with a Poisson manifold $\mathcal M$ and we introduce a parameter $\theta$ of dimension $\textrm{length}^2$ as
\be \{x^\mu,x^\nu\}=\theta\,\theta_0^{\mu\nu}\ .\ee
To quantize this manifold, we require a quantization map
\be
\I\ : \ \mathcal C(\mathcal M)\to \A
\ee
from the algebra of function $\mathcal C(\mathcal M)$ to some (perhaps infinite dimensional) matrix algebra $\A$, with the following properties
\begin{align}
    \I(fg)-\I(f)\I(g)\,\to\,&0\nonumber,\\
\frac{1}{\theta}\big[\I(i\{f,g\})-[\I(f),\I(g)]\big]\,\to\,&0\ ,\nonumber\\
&\textrm{as }\theta\,\to\, 0\ ,\ \forall\,f,g\in \mathcal C(\mathcal M)\ .
\end{align}
Using this map, we can define a new product on $C(\mathcal M)$ denoted by $\ast$ as
\be f\ast g=\I^{-1}\lr{\I(f)I(g)}\ .\ee
If we have the manifold $\mathcal M$ embedded in $\mathbb R^d$ by coordinate functions $x^a$, we can use the images of these as fuzzy coordinates
\be X^a=\I(x^a)\ .\ee
Using these, one can construct geometric objects defined on $\mathcal M$ also on the fuzzy version living in the algebra $\A$.

\subsection{Field theory on fuzzy spaces}

As we have seen in the previous section, real functions on fuzzy spaces are given by Hermitian matrices. These will therefore be the scalar fields and we can straightforwardly write an action for the scalar field theory by translating the relevant objects to the fuzzy setting \cite{sF22}. From now on, we will consider the case of the fuzzy sphere, where the action is as follows
\be\label{fuzzyaction}
S(M)=
	\frac{4\pi R^2}{N}\textrm{Tr}\slr{
	\half
	M\frac{1}{R^2} [L_i,[L_i,M]]
	+\half m^2 M^2
	+V(M)}\ .
\ee
For simplicity we set the radius of the sphere to $1$, we will concentrate on the quartic case $V(M)=gM^4$ and we rescale the quantities to hide the volume prefactor.

Since our fields now have finite number of degrees of freedom, the path integral over all field configuration is just a well-defined matrix integral and we can define quantum theory of the scalar field via the correlation functions
\be
\left\langle F\right\rangle=\frac{\int M\, F(M)e^{-S(M)}}{\int dM\,e^{-S(M)}}\ .
\ee
To evaluate such correlation functions, we invoke the standard machinery of Feynman diagrams.

We first expand the matrix $M$ in the basis of polarization tensors $T_{lm}$ as
\be M=\sum_{l=0}^{N-1}\sum_{m=-l}^l c_{lm}T_{lm}\ .\ee
The polarization tensors are eigenmatrices of the matrix Laplacian \cite{varch}
\be [L_i,[L_i,T_{lm}]]=l(l+1)T_{lm}\ \ ,\ \ [L_3,T_{lm}]=mT_{lm}\ \ , \ \ T_{lm}^\dagger=(-1)^m T_{l-m}\ ,\ee
thus for real field $\bar c_{lm}=(-1)^m c_{l-m}$. They are orthogonal
\be \trl{T_{lm}^\dagger T_{l'm'}}=\delta_{ll'}\delta_{mm'}\ee
and therefore diagonalize the quadratic part of the action. This means that we get the following contraction between the momentum modes
\be \contraction{}{\bar c}{{}_{lm}}{\bar c}\bar c_{lm}c_{l'm'}=\frac{1}{l(l+1)+m^2}\delta_{ll'}\delta_{mm'}\ . \ee
So far very little is truly different from the commutative case. However, the real differences start to emerge when we consider the interaction part.

The momentum space vertex is given by
\be V(l_1m_1,l_2m_2,l_3m_3,l_4m_4)=g\trl{T_{l_1m_1}T_{l_2m_2}T_{l_3m_3}T_{l_4m_4}}\ee
and we clearly see where the issue is going to be. The vertex factor is not invariant under any permutation of the external momenta; only cyclic changes leave it unchanged. This means that we will need to be very careful with the way legs attach to vertices when computing Feynman diagrams.

\subsubsection{UV/IR mixing on the fuzzy sphere}

\begin{figure}
    \centering
    \includegraphics[width=0.5\textwidth]{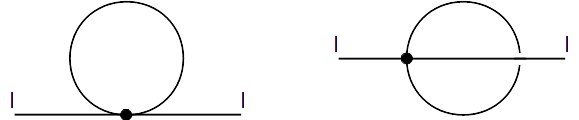}
    \caption{The standard planar diagram of $\phi^4$ on the left and non-planar diagram, different from the diagrams in the commutative theory, on the right.}
    \label{1loopdiags}
\end{figure}

We would now like to compute the one-loop two point function of (\ref{fuzzyaction}). In the usual theory there would be just one diagram to evaluate, the left diagram in the figure \ref{1loopdiags}. But since we are alert about the way legs contract in the vertices we realize that the second diagram in the same figure might turn out to be different and needs careful analysis. The defining difference between the two graphs is their planarity. The left diagram is planar, opposing to the second one, which cannot be drawn on a plane and needs topologically more complicated surface to be drawn without self-intersections.

We refer the reader to the original work \cite{uvir2} or one of the excellent reviews \cite{noncom1,noncom2,bal,lecturesydri} and only present the final formulas for the planar and non-planar diagram
\begin{align*}
I^P=&\sum_{j=0}^{N-1} \frac{2j+1}{j(j+1)+m^2}\ ,\\
I^{NP}=&	\sum_{j=0}^{N-1} \frac{2j+1}{j(j+1)+m^2}N(-1)^{l+j+N-1}
\left\lbrace\begin{array}{lll} l & s & s\\ j & s & s \end{array}\right\rbrace\ ,\ s=\frac{N-1}{2}\ .
\end{align*}
The planar contribution is the usual momentum loop summation that one encounters in the commutative field theory. The non-planar contribution is very different. The object with six entries in curly brackets is a $6j$-symbol, group theoretical factor stemming from the traces of the polarization tensors \cite{varch}. The loop summation depends on the incoming momentum $l$.

The most interesting thing happens when we try to take the commutative limit $N\to\infty$. The non-planar contribution does not become equal to the planar one, but keeps a finite difference equal to \cite{uvir2}
\be I^P-I^{NP}\to 2\sum_{k=1}^l \frac{1}{k}\ .\ee
The commutative limit of one-loop effective action of the theory (\ref{fuzzyaction}) is different from one-loop effective action of the commutative theory by the above contribution. This phenomenon is called the UV/IR-mixing.

In non-compact noncommutative spaces it is a full blown mixing of UV and IR divergences of the theory \cite{uvir1}. The physics at very small scales has huge consequences at large distances, the separation of scales needed for renormalization of the theory is destroyed and this does not change even in the commutative limit. In our fuzzy sphere setting the situation is little milder, as the theory is finite at finite $N$ and no additional regularization is needed. Still, the difference between the planar and non-planar diagrams introduces a non-local contribution to the effective action and is sometimes called the noncommutative anomaly.

\subsubsection{Phase structure on the fuzzy sphere}

Probably the best studied property of fuzzy scalar field theories is their spontaneous symmetry breaking patters, or phase structure.

The two dimensional euclidean $\phi^4$ field theories have two different phases \cite{comr2}. A disorder phase, where the field oscillates around zero, $\avg \phi=0$. If the wells of the potential are too deep, the quantum fluctuations are not large enough to keep the field out of the well and it oscillates around a nonzero value $\avg \phi=\phi_0$, which is one of the true minima of the potential. This uniform order phase spontaneously breaks the $\phi\to-\phi$ symmetry of the potential. Later, numerical simulations have been able to identify the transition line between these two phases \cite{comr2num}.

Soon after the discovery of the UV/IR-mixing it has been realized that one of the consequences of this phenomenon is existence of a third phase of fuzzy field theories \cite{NCphase1,NCphase2}. This phase, the non-uniform order or the striped phase, breaks the translation invariance of the underlying space, as the field does not oscillate around a single value in the whole space.

This phenomenon has been observed numerically in large body of work: for the fuzzy sphere \cite{OConSamo,num09,num14}, fuzzy disc \cite{num_disc}, fuzzy sphere with a commutative time \cite{num_RSF2} and fuzzy torus \cite{num14panero2}. See \cite{panero15} for a review of this topic. Analytical approaches to the problem \cite{OConSaman,samann,samann2,samanRfuzzyS,samanfuzzydisc,poly13,steinacker05,PNT12,jt18,jt15b,msjt20} have been centered around the matrix models description of the fuzzy field theory we will describe in the section \ref{sec4}.

\begin{figure}
    \centering
    \includegraphics[width=0.35\textwidth]{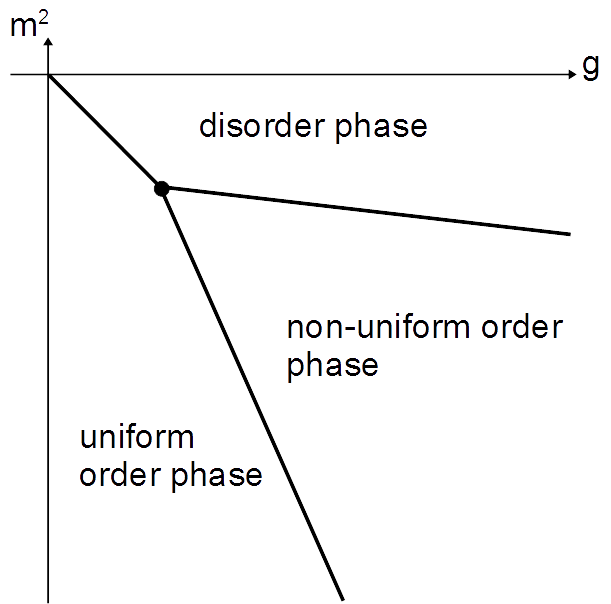}
    \caption{Schematic phase diagram of a generic fuzzy field theory as found by various numerical works.}
    \label{schemdiag}
\end{figure}

Universally the following parameter space properties of the fuzzy field theories have been identified. There are three transition lines between the phases described above and schematically shown in the figure \ref{schemdiag}. All three lines meet at one point, the triple point of the theory. For the fuzzy sphere the location of the triple point has been found in \cite{OConSamo} to be
\be r_c=-0.054\pm0.003\ ,\ g_c=0.005\pm0.00025\ .\label{samotriple}\ee

\subsubsection{Other properties of field theories on the fuzzy sphere}\label{sec_other}

Recently, two other very interesting properties of fuzzy field theories have been investigated numerically. In both cases, the results show different behavior from the commutative counterparts.

\paragraph{Correlation functions.} In \cite{correlationFunctions0,correlationFunctions1,correlationFunctions2}, authors have investigated properties of correlation functions $\avg{\phi \phi}$ in the theory (\ref{fuzzyaction}), most notably their change as we cross from the disorder to uniform order phase (described in the previous section), which is the standard phase transition line present also in the commutative field theory. Surprisingly, authors find that at the phase transition, the correlation functions behave differently from the commutative case and at short distances seem to agree with the tricritical Ising model. 

\paragraph{Entanglement entropy.} In local theories, the entanglement entropy related to a given region is famously proportional to the surface of the boundary of the region. This changes in non-local theories and in several numerical works \cite{entanglement1,entanglement2,entanglement3} authors have investigated properties of the entanglement entropy in fuzzy scalar field theories. For free fields they have been able to identify a volume rather than surface law, as expected due to the non-locality of the theory.

\section{Matrix models of fuzzy field theories}\label{sec4}

We are finally ready to set up the analysis of the fuzzy field theories using matrix models. After we cover the construction of the relevant matrix models, we will describe the most recent results in description of the phase structure and mention challenges one needs to meet when dealing with problems beyond.

\subsection{Kinetic term effective action}

As we have seen, the questions of the fuzzy field theory are answered in terms of the correlation functions
\be
\avg{F}=\frac{1}{Z}\int \lr{\prod d\lambda_i}\,e^{-N^2\slr{\frac{1}{N}\sum_i\lr{r\lambda_i +V(\lambda_i)} -\frac{2}{N^2}\sum_{i<j}\log|\lambda_i-\lambda_j|}}\int dU\,F(\Lambda,U)e^{-N^2\half\trl{M\K M}}\ 
\ee
where we have switched from the standard field theory notation $m^2$ to the matrix models notation for the quadratic coefficient $r$. If we ask $SU(N)$ invariant questions $F(\Lambda)$, the $dU$ integral can be done once and for all and depends only on the eigenvalues $\lambda_i$ of the matrix $M$.

We thus define the kinetic term effective action as
\be\label{31anfintegral}
\int dU\,e^{-N^2\half\trl{M\K M}}=:e^{-N^2 S_{eff}(\Lambda)}\ .
\ee
As we will see $S_{eff}$ will turn out to be a rather complicated multi-trace action. A complete result is not available at the moment and we only have two, rather different, approximations.

\subsubsection{Perturbative angular integral}
The first approach is to evaluate the integral (\ref{31anfintegral}) perturbatively in powers of the kinetic term, where the $dU$ integrals are tedious, but manageable for first few orders. This was first done in \cite{OConSaman} for the fuzzy sphere and later refined and extended in \cite{samann} and \cite{samann2}. We will not describe all the steps here and refer reader to the original papers.

We only give the final large $N$ result for the fuzzy sphere from \cite{samann2} valid up to the eight order in eigenvalues
\be
S_{eff}=\Bigg[\half \lr{\half t_2-\frac{1}{24}t_2^2+\frac{1}{2880}t_2^4}-\frac{1}{432}t_3^2-\frac{1}{3456}\lr{t_4-2t_2^2}^2\Bigg]\ ,\label{32rearr}
\ee
where
\be
t_n=\frac{1}{N}\trl{M-\frac{1}{N}\trl{M}}^n\ .
\ee
We see that the effective action is function of the symmetrized moments $t_n$. Similar result is available for the fuzzy disc \cite{samanfuzzydisc}.

\subsubsection{Nonperturbative angular integral approximation}

A different approximation to the kinetic term effective action (\ref{31anfintegral}) can be derived from a result first obtained in \cite{steinacker05} and later derived using different methods in \cite{PNT12}.

Consider the free field theory given by the action
\be
S(M)=\half\trl{M\K M}+\half r \trl{M^2}\ .\label{4.2}
\ee
Using diagrammatical analysis of the corresponding matrix model one can show that the large $N$ solution of this model is again a semicircle distribution with a modified radius
\be\label{31radscaled}
R=2\sqrt{f(r)}\ ,
\ee
where the function $f(r)$ is given by
\be
f(r)=\int_0^1dx\frac{2x}{x^2+r}=\log\lr{1+\frac{1}{r}}\ .\label{31efforSF}
\ee
This means that the effective action (\ref{31anfintegral}) can only be given by a second moment part, which produces the rescaled semicircle, and a part that vanishes when evaluated on the semicircle distribution \cite{poly13}. Explicitly
\be\label{32sefffull}
S_{eff}=\half F(t_2)+a_1 t_3^2+(b_1+b_2 t_2)(t_4-2t_2^2)^2+c_1(t_6-5 t_2^3)(t_4-2 t_2^2)+\ldots=\half F(c_2)+\mathcal R
\ee
for some function $F(t_2)$ to be determined by the condition (\ref{31radscaled}) and a remainder term $\mathcal R$ which this approach cannot capture. The remainder term is a function of the following combinations of the symmetrized moments\footnote{Technically one needs product of at least two such term, since they need to vanish in the saddle point condition.}
\be\label{32difsemi}
t_{2n}-C_n t_2^n=t_{2n}-\frac{(2n)!}{n!(n+1)!} t_2^n\ ,\ t_{2n+1}\ ,
\ee
which vanish for the semicircle and are written in increasing powers of the eigenvalues.  

Using the know radius for the semicircle distribution and its second moment
\be
R^2=\frac{4}{r_{eff}}=\frac{4}{r+F'(t_2)}\ ,\ t_2=\frac{1}{r_{eff}}=\frac{1}{r+F'(t_2)}\ ,
\ee
we obtain a condition for the function $F$ as follows
\be
F'(x)=\frac{1}{x}-f^{-1}(x)\ .
\ee
Using (\ref{31efforSF}) this yields, with a condition $F(0)=0$,
\be
F_{S^2_F}(t_2)=\log\lr{\frac{t_2}{1-e^{-t_2}}}=\half t_2 - \frac{1}{24}t_2^2 + \frac{1}{2880}t_2^4+\ldots\ .\label{Fsphere}
\ee
Notice, how the expansion of this function matches the first part of (\ref{32rearr}), while the second part of that expression is exactly of the form expected in (\ref{32sefffull}).

\subsection{Second moment models}\label{sec4.2}

As one approximation to the fuzzy field theory matrix models on the fuzzy sphere one can drop the remainder term $\mathcal R$ in (\ref{32sefffull}) and take the form of the function $F$ (\ref{Fsphere}). In this section, we will give results of the analysis of such multi-trace matrix models and the resulting phase structure. We will provide also some intermediate steps, with more details in the original works \cite{jt18,msjt20}. 

When viewed as a multi-trace matrix model of the section \ref{sec2multi} this is a particular case of (\ref{24genmlti}) with
\be f(c_2,c_1)=\half F\lr{c_2-c_1^2} \ .\label{5sphereF}\ee

Expression for the symmetric regime $c_1=0$ of such model have been given in the section \ref{sec2multi}.

Deriving the relevant condition for the asymmetric regime (\ref{ass1c}) is tedious, but straightforwardly follows the symmetric case and yields
\be
4\frac{4+15\delta^2 g + 2 r \delta}{\delta(4+9\delta^2 g)}-F'\lr{\frac{\delta \lr{64  + 160 \delta^2 g+144 \delta^4 g^2+81 \delta^6 g^3+36 \delta^3 g r + 27 \delta^5 g^2 r}}{64(4+9\delta^2 g)}}=0\ ,\label{4_cond3}
\ee
with the other parameter of the distribution given by
\be
D^2=\frac{-r(4 +3 \delta^2 g )-12 \delta g-9 \delta^3 g^2}{4 g \left(4+9 \delta^2 g\right)}
\ee
and the free energy given by
\begin{align}
\F_{AS1C}\,=\,&-\frac{1}{128 g \left(4+9 \delta^2 g\right)^2}\Big[
6075 \delta^8 g^5
+3240 \delta^6 g^4 (4+r \delta)
+144 \delta^4 g^3 \left(29+40 r \delta+3 \delta^2 r^2\right)\no
+8 \delta^2 &g^2 \left(-144+352 r \delta+117 \delta^2 r^2\right)+64 g \left(-8+8 r \delta+9 \delta^2 r^2\right)
+128 r^2\Big]+\no
+\half F&\lr{\frac{\delta \lr{64  + 160 \delta^2 g+144 \delta^4 g^2+81 \delta^6 g^3+36 \delta^3 g r + 27 \delta^5 g^2 r}}{64(4+9\delta^2 g)}}-\half \log\lr{\frac{\delta}{4}}\ .\label{4_free1AC}
\end{align}

\paragraph{Symmetric phase transition.}

The matrix phase transition between the symmetric one-cut and two-cut solution is obtained directly using (\ref{multimatrixtrafo}) and yields
\be\label{5_line1}
r=-4\sqrt{g}-F'(1/\sqrt{g})=-5\sqrt{g}-\frac{1}{1-e^{1/\sqrt{g}}}\ .
\ee
We will drop the exponentially suppressed terms, as they will be negligible in the relevant region of parameter space.

\paragraph{Asymmetric one-cut to symmetric two-cut phase transition.}

The equations (\ref{fconditon},\ref{fconditon2},\ref{4_cond3}) are clearly out of reach of any direct analytical treatment. They have been analyzed numerically in \cite{jt18}. In \cite{msjt20} analytical solution based on perturbative solution and subsequent Pade approximation has been found and we will present the results here.

Equations (\ref{fconditon2}) and (\ref{4_cond3}) allow for a nice expansion in powers of $1/r$ in the case of a large and negative $r$. The large and small expansion of $F(t_2)$ is needed in each case respectively. One then solves the defining conditions order by order in $1/r$ and obtains the free energies of the asymmetric one-cut and the symmetric two-cut solutions. We drop the exponentially suppressed terms in the large $t_2$ expansion of $F(t_2)$ for the two-cut solution, as they give a negligible contribution.

To find the phase transition between the two we solve the condition $\F_{AS1C}=\F_{S2C}$, again order by order in powers of $1/r$ and obtain the following result
\be\label{3trafo_pert}
g=\frac{1}{16 e^{3/2}}+\frac{1}{32 e^{3/2} r}+\frac{9+5 e^{3/2}}{384 e^3 r^2}+\frac{141+16 e^{3/2}}{3072 e^3 r^3}+\frac{13545+18240 e^{3/2}+764 e^3}{368640 e^{9/2} r^4}+\ldots\ .
\ee
To obtain a non-perturbative transition line we treat the above expression to a Pade approximation in $1/r$ and obtain
\be\label{padetrafo}
g=\frac{\frac{1}{16 e^{3/2}}
+\frac{4095+4890 e^{3/2}+4 e^3}{640 e^{3/2} \left(162-243 e^{3/2}+2 e^3\right) r}+\frac{-92745+43920 e^{3/2}+4164 e^3+4 e^{9/2}}{7680 e^3 \left(162-243 e^{3/2}+2 e^3\right) r^2}+\ldots
}{1-\frac{3 \left(-285-3250 e^{3/2}+12 e^3\right)}{40 \left(162-243 e^{3/2}+2 e^3\right) r}+\frac{-121905+66330 e^{3/2}-30396 e^3+20 e^{9/2}}{480 e^{3/2} \left(162-243 e^{3/2}+2 e^3\right) r^2}+\ldots}\ .
\ee
It is not difficult to check that this phase transition is of the first order, since the first derivative of the free energy is discontinuous as one crosses the phase transition line.

\paragraph{Asymmetric one-cut to symmetric one-cut phase transition.}

This transition is the most involved to find. The reason is that the condition (\ref{fconditon}) does not directly allow for any nice large or small $r$ treatment. What can be done is an expansion around the phase transition value (\ref{5_line1}). What follows is quite a technical endeavor with an unilluminating result. Let us refer the curious reader to \cite{msjt20} and let us proceed to more interesting aspects of the matrix model.

This phase transition is also of the first order and is often referred to as the Ising transition, since it describes a transition from disordered phase with $c_1=0$ to an uniformly ordered phase with $c_1\neq 0$.

\paragraph{Phase diagram and the triple point.}

\begin{figure}
    \centering
    \includegraphics[width=0.8\textwidth]{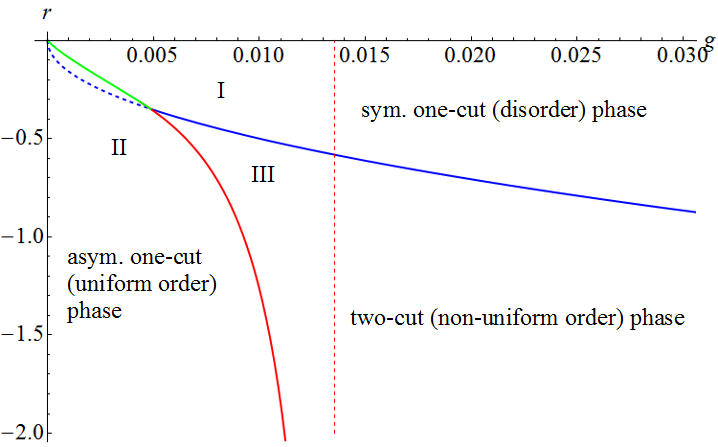}
    \caption{The phase diagram of the model with the effective action (\ref{5sphereF}). The green line separates the symmetric one-cut solution from the asymmetric one-cut solution and is obtained using $1/r^{12}$ expansion, the blue line separates the symmetric one-cut and  the two-cut solutions and is exact, the red line separates the two-cut solution from the asymmetric one-cut solution and is computed using $1/r^{20}$ expansion in (\ref{padetrafo}). The vertical dashed line denotes the asymptote of the red line at ${g=1/16e^{3/2}}$.}
    \label{phasediag2}
\end{figure}

All three transition lines are shown in the figure \ref{phasediag2}. The general features of the diagram are as expected. Most importantly there is a region of a preferred asymmetric one-cut solution. The three transition lines meet at the triple point and the value of the critical coupling is
\be g_c=0.0048655\ ,\ee
obtained numerically as an intersection of the lines (\ref{5_line1}) and (\ref{padetrafo}). This value is reasonably close the value (\ref{samotriple}) obtained in \cite{OConSamo}.

However, there are features of the diagram which are not completely satisfactory. One of them is the value of the critical quadratic parameter $r$, which is too large roughly by a factor of $6$. Moreover, the asymmetric one-cut to symmetric two-cut transition line (\ref{padetrafo}) behaves differently for large values of $r$. A linear line has been found in previous numerical studies, while the line (\ref{padetrafo}) converges to ${g=1/16e^{3/2}}$.

We expect all these differences to be a consequence of our approximation, when we neglected the remainder term $\mathcal R$ in (\ref{32sefffull}). And to correct them we need to find a way how to incorporate these higher moment terms.

\subsection{Beyond second moment models}\label{sec4.3}

The next natural step is to analyze the second available approximation to the kinetic term effective action (\ref{32rearr}). This is rather straightforward to do following the approach of the section \ref{sec2multi}. The results are however not very useful, at least close to the origin of the parameter space where the triple point is located \cite{jt15b}.

For situations when the moments of the distributions are small and expansion (\ref{32rearr}) works well one can do the same large and negative $r$ analysis as we did in the previous section. However for the two-cut solution, where the moments are large, this approximation is not good and does not yield any useful solutions at large $r$, which could be translated towards small values of parameters as in the previous section.

Moreover, numerical analysis of the model shows that close to the origin of the parameter space, where the potential is weak and the eigenvalues are spread out, the model behaves very differently from what we would expect from fuzzy field theory.

The root of this problem is clear. Comparing the formulas (\ref{Fsphere}) and (\ref{32rearr}) we see that the expansions in (\ref{32rearr}) are the first few terms in an alternating series which describes a function that probably grows much slower as was the case for the $t_2$ expansion. Cutting the expansion and taking it at face value leads quickly to issues for larger values of the moments.

In the picture of the particle dynamics, the effective action (\ref{32rearr}) introduces an interaction, which can be repulsive for certain configurations and destabilizes the whole configuration.

\section{Challenges and outlook}

We have reviewed the description of fuzzy field theory on the fuzzy sphere by certain Hermitian matrix models. These models are characterized by the kinetic term effective action in the probability distribution, which makes the models far more complicated than the standard cases.

We have shown that a non-perturbative approximation to this action leads to a model, which describes some of the features of the phase diagram of the theory well. However, is not sufficient to explain all its properties. We conclude this review with further questions that can be naturally asked.

\paragraph{Beyond the second moment.} One of them has been mentioned in the section \ref{sec4.3}. We have some information about the behavior of the kinetic term effective action beyond the second moment (\ref{32rearr}) from perturbative calculations \cite{samann2}. We however need to find a way to include these terms in a non-perturbative fashion \cite{MSJTcorfu19b}.

\paragraph{Beyond the fuzzy sphere.} There are numerical \cite{num_disc,num_RSF2,num14panero2} and perturbative analytical \cite{samann,samanRfuzzyS,samanfuzzydisc} results for properties of field theories on spaces beyond the fuzzy sphere. The second moment method described in the section \ref{sec4.2} can be generalized to different spaces by considering a different kinetic term of the theory \cite{poly13,msjt20}. It would be very interesting to see what kinds of results we could obtain for spaces such as fuzzy disc, fuzzy torus, fuzzy $\mathbb C P^2$, etc.

\paragraph{Kinetic term effective action.} We have limited information about the kinetic term effective action (\ref{31anfintegral}). It is clearly of importance to either compute the angular integral completely or to get more information following further the calculation of \cite{samann,samann2}. Also, great progress in understanding of related matrix models has been reported in \cite{gw19} and these results should be relevant in this setting too.

\paragraph{UV/IR-mixing free theories.} Naive generalizations of field theories to fuzzy setting are plagued with UV/IR-mixing and are very different from commutative theories. In the case of non-compact spaces, this phenomenon ruins renormalization of the theory completely. It is therefore necessary to consider modified version of the theories, which are UV/IR-mixing free. It is conjectured that phase diagram of such theories should have no non-uniform order, or two-cut, phase. Example of a modification is available for $\phi^4$ on the fuzzy sphere \cite{DolOConPres} and the first results in the analysis of the corresponding matrix models are given in \cite{msjt20}.

\paragraph{Other properties of fuzzy field theories.} As mentioned in the section \ref{sec_other}, numerical studies have investigated properties of fuzzy field theories beyond the phase structure. It would be very interesting to see if the models described in this review, or any further generalizations, capture features like two-point correlation functions or entanglement entropy. Such investigations could supplement the numerical results and provide an analytical handle on fuzzy field theories. Finally, this could give us further clues on where to proceed with the analytical treatment of fuzzy field motivated matrix models.

\acknowledgments

This work was supported by \emph{VEGA 1/0703/20} grant. The work of JT was supported by \emph{Alumni FMFI} foundation as a part of the \emph{N\'{a}vrat teoretikov} project, the work of M\v S has been supported by \emph{UK/432/2019} and \emph{UK/432/2018} grants. Both of the authors have greatly benefited from the support of the COST MP-1405 action \emph{Quantum structure of spacetime}.

\end{document}